\documentclass{iau}
\usepackage{graphicx}

\newcommand{\ave}[1]{\left\langle #1\right\rangle}
\newcommand{\simgt}{\lower.5ex\hbox{$\; \buildrel > \over \sim \;$}}
\newcommand{\simlt}{\lower.5ex\hbox{$\; \buildrel < \over \sim \;$}}

\newcommand{\bx}{{\bf x}}
\newcommand{\bk}{{\bf k}}
\newcommand{\btheta}{{\boldsymbol{\theta}}}
\newcommand{\bq}{{\bf q}}
\newcommand{\bl}{{\bf l}}

\newcommand{\tdelta}{\tilde{\delta}}
\newcommand{\tW}{\tilde{W}}

\title[Statistical challenges in WL cosmology] %% give here short title %%
{Statistical challenges in weak lensing cosmology}

\author[M. Takada]
{Masahiro Takada}

\affiliation{Kavli Institute for the Physics and Mathematics of the Universe
(WPI),
Todai Institutes for Advanced Study
The University of Tokyo, Chiba 277-8583, Japan \\email:{\tt masahiro.takada@ipmu.jp}}

\pubyear{2014}
\volume{xxx}  %% insert here IAU Symposium No.
\pagerange{119--126}
\jname{Statistical Challenges in 21st Century Cosmology}
\editors{A.C. Editor, B.D. Editor \& C.E. Editor, eds.}
\begin{document}

\maketitle

\begin{abstract}
Cosmological weak lensing is the powerful probe of cosmology.
%method for constraining 
%the
%nature of dark energy or testing gravity theory on cosmological
%scales. 
Here we address one of the most fundamental, statistical
questions inherent in weak lensing cosmology: whether or not we can {\em
recover} the initial Gaussian information content of large-scale structure by combining the weak lensing
observables, here focused on the weak lensing power spectrum and
bispectrum.
To address this question we fully take into account correlations between
the power spectra of different multipoles and the bispectra of different
triangle configurations, measured from a finite area survey. In
particular we show that super-survey modes whose length scale is larger
than or comparable with the survey size cause significant sample
variance in the weak lensing correlations via the mode-coupling with
sub-survey modes due to nonlinear gravitational clustering -- the
so-called {\em super-sample variance}.  In this paper we discuss the
origin of the super-sample variance and then study the information
content inherent in the weak lensing correlation functions up to
three-point level.  
\end{abstract}

\firstsection % if your document starts with a section,
              % remove some space above using this command.
\section{Introduction}
Cosmological weak lensing is one of the most powerful cosmological
probes, as it allows us to directly map out the distribution of matter
in the universe without assumptions about galaxy biases (see Heymans
\etal~2013; More \etal~2014 for the recent results). Upcoming galaxy
surveys such as the Subaru Hyper Suprime-Cam Survey (\cite[Takada
2010]{Takada:10}) aim to use the high-precision weak lensing
measurements to tackle questions on fundamental physics including the
origin of cosmic acceleration and neutrino masses.

Most of the useful weak lensing signals are in the nonlinear clustering
regime, over a range of multipoles around $\ell\sim $a few
thousands. Due to the mode-coupling nature of nonlinear structure
formation, the weak lensing field at the scales of interest display
prominent non-Gaussian features. Thus the two-point correlation function
or the Fourier-transformed counterpart, the power spectrum, no longer
fully describes the statistical properties of the weak lensing
field. Which statistical quantities or their combination can be optimal
to extract a maximum information of the weak lensing field is still an
open question and needs to be carefully explored in order to attain the
full potential of the weak lensing surveys. Although weak lensing
cosmology involves various statistical issues such as an accurate
measurement of galaxy shapes, astronomical data reduction, and parameter
estimation, in this paper we focus on the above statistical question.

\section{Weak lensing cosmology}

The weak lensing convergence field is expressed as a weighted projection
of the three-dimensional mass density fluctuation field along the line
of sight. For a source galaxy at the radial distance $\chi_s$ and in the
angular direction $\btheta$, the convergence field is given by
\begin{equation}
 \kappa(\btheta)=\frac{3}{2}\Omega_{\rm m0}H_0^2\int^{\chi_s}_0\!d\chi~
a^{-1}\chi\left(1-\frac{\chi}{\chi_s}\right)\delta_m(\chi,\btheta),
\end{equation}
where $\delta_m$ is the mass density fluctuation field along the line of
sight and we assumed a flat geometry universe. Although the weak lensing
is observationaly estimated from the ellipticities of source galaxy
shapes, the so-called weak lensing shear field, the shear field is
equivalent to the convergence field in the weak lensing regime, so we
throughout this paper work on the convergence field.  As obvious from
the above equation, the statistical properties of the weak lensing field
reflect those of the underlying mass density. If the mass density field
is a Gaussian random field, which is the case in the linear regime, the
weak lensing field is also Gaussian. If the mass field is non-Gaussian,
which is the case in the nonlinear regime, the weak lensing should
inevitably display non-Gaussian features.

The weak lensing field is measurable only in a statistical way. The most
conventional method used in the literature is the two-point correlation
function. Using the Limber's approximation and the flat-sky
approximation, the Fourier-transformed counterpart, the power spectrum
is given as
\begin{equation}
 P_\kappa(\ell)=\int^{\chi_s}_0\!d\chi~W_{\rm GL}(\chi)^2\chi^{-2}P_\delta\left(k=\frac{l}{\chi};\chi\right), 
\label{eq:pkappa}
\end{equation}
where we defined the lensing efficiency function $W_{\rm GL}(\chi)\equiv
(3/2)\Omega_{\rm m0}H_0^2a^{-1}\chi(1-\chi/\chi_s)$, and $P_\delta(k;a)$
is the mass power spectrum at redshift $a=a(\chi)$. Similarly, the
lensing bispectrum, which is the lowest-order correlation function to
measure the non-Gaussianity, is defined as
\begin{equation}
 B_\kappa(\bl_1,\bl_2,\bl_3)=\int^{\chi_s}_0\!d\chi~W_{\rm GL}(\chi)^3\chi^{-4}
B_\delta(\bk_1,\bk_2,\bk_3;\chi),
\label{eq:bkappa}
\end{equation}
where $\bk_i=\bl_i/\chi$ and $B_\delta(\bk_i)$ is the mass bispectrum,
and a set of the three wavevectors satisfies the triangle condition in
Fourier space: e.g., $\bl_1+\bl_2+\bl_3={\bf 0}$.  While the power
spectrum is a one-dimensional function of the wavelength $l$, the
bispectrum is given as a function of triangle configurations.  Likewise
the $n$-point correlation function of the weak lensing field arises from
the $n$-point function of the mass density field.

The weak lensing correlation functions are sensitive to both the
geometry of the Universe and the growth of matter clustering via the
lensing efficiency function and the mass correlation functions. With
these dependences weak lensing cosmology is expected to be one of the
most powerful probes for constraining cosmological parameters as well as
testing theory of gravity on cosmological scales (\cite[Takada \& Jain
2004; Oguri \& Takada 2011]{TakadaJain:04,OguriTakada:11}).

\section{Statistical power of weak lensing correlation functions}

\subsection{Super-sample covariance}

In order to realize the constraining power of the weak lensing
correlation functions for a given survey, we need to quantify
statistical uncertainties in the measured correlation functions.  The
important source of the statistical uncertainties is the {\em sample
variance} arising due to a finite number of Fourier modes sampled from a
finite survey volume.  Recently we developed a simple, unified approach
to describing the impact of super-sample covariance, which arises from
modes that are larger than or comparable with the survey size, on the
correlation functions. The method is written in a general
form and can be applied to any large-scale structure probe.  In this
section we briefly review the theory following \cite{TakadaHu:13} (also
see Hamilton et al. 2006 for the pioneer work).

Since the statistical properties of the weak lensing field reflect those
of the mass density field as we discussed above, we here consider the
power spectrum of the three-dimensional mass density field. For a finite
volume survey, the observed field is generally expressed as
\begin{equation}
 \delta_{m,W}(\bx)=\delta_m(\bx)W(\bx),
\end{equation}
where $W(\bx)$ is a survey window function; $W(\bx)=1$ if $\bx$ is in
the survey region, otherwise $W(\bx)=0$.  For a finite-area weak lensing
survey, one can think of the windowed mass density field as the mass
density field in the finite volume around a certain lens redshift and
confined with the survey area.  The Fourier-transformed density field is
given as $\tdelta_{m,W}(\bk)=\int\!d^3\bq/(2\pi^3)\tW(\bk-\bq)
\tdelta_{m}(\bq)$. Through the window function that has support for
$q\simlt 1/L$ where $L=V^{1/3}$ is the typical size of the survey, we
can properly take into account the effects of super-survey modes that
are comparable with or larger than the survey size.

Then we can define the power spectrum
estimator as
\begin{equation}
 \hat{P}_\delta(k_i)\equiv \frac{1}{V_W}\int_{|\bk|\in k_i}\!\!\frac{d^3\bk}{V_{k_i}}\tdelta_{m,W}(\bk)\tdelta_{m,W}(-\bk),
\end{equation}
where the integral is over a shell in $k$-space of width $\Delta k$,
volume $V_{k_i}\simeq 4\pi k_i^2\Delta k$ for $\Delta k/k_i\ll 1$, and
the effective survey volume is defined as $V_W=\int\!d^3\bx W(\bx)$.
The ensemble average of its estimator is a convolution of the underlying
power spectrum with the window
\begin{equation}
 \ave{\hat{P}_\delta(k_i)}=\int_{|\bk|\in k_i}\!\!\frac{d^3\bk}{V_{k_i}}
\int\!\frac{d^3\bq}{(2\pi)^3}|\tW(\bq)|^2P_\delta(\bk-\bq),
\end{equation}
%
%The survey window has support for $q\simlt 1/L$ where $L=V^{1/3}$ is the
%typical size of the survey. 
Thus for $k\sim 1/L$ this estimator is
biased low compared to the true power spectrum due to transfer of power
into the fluctuation in the spatially-averaged density of the survey
volume. For $k\gg 1/L$ this bias becomes progressively smaller 
since the underlying power spectrum is expected to be smooth across
$\Delta k \sim 1/L$.

The covariance matrix describes statistical uncertainties in the power
spectrum estimation, defined as
\begin{eqnarray}
 C^P_{ij}\equiv {\rm Cov}[P_\delta(k_i),P_\delta(k_j)]&=&
\ave{\hat{P}_\delta(k_i)\hat{P}_\delta(k_j)}
-\ave{\hat{P}_\delta(k_i)}\ave{\hat{P}_\delta(k_j)}\nonumber\\
&\simeq&C^G_{ij}+\frac{1}{V_W}\bar{T}_W(k_i,j_j).
\label{eq:cov_def}
\end{eqnarray}
Here the Gaussian piece is
\begin{equation}
 C^G_{ij}=\frac{1}{V_W}\frac{(2\pi)^3}{V_{k_i}}2P_\delta(k_i)^2\delta^K_{ij}, 
\end{equation}
with $\delta^K_{ij}=1$ if $k_i=k_j$ to within the bin width, otherwise
$\delta^K_{ij}=0$. Here $V_{k_i}/[(2\pi)^3/V_W]$ is the number of
Fourier modes used in the power spectrum estimation at the bin
$k_i$. The prefactor ``$2$'' in $C^G_{ij}$ arises from the fact that the
density field is real, yielding $\tdelta_{m}(\bk)=\tdelta^\ast(-\bk)$,
and therefore the modes of $\bk$ and $-\bk$ are not independent.  The
Gaussian piece has only the diagonal components, ensuring that the power
spectra of different bins are uncorrelated with each other. The second
term, proportional to $\bar{T}_W(k_i,k_j)$, is the non-Gaussian
contribution arising from the connected 4-point function or trispectrum,
convolved with the survey window function:
\begin{eqnarray}
 \bar{T}_{\delta,W}(k_i,k_j)&=&\frac{1}{V_W}\int_{|\bk|\in
  k_i}\!\!\frac{d^3\bk}{V_{k_i}}
\int_{|\bk|\in k_j}\!\!\frac{d^3\bk'}{V_{k_j}}
\int\!\left[\prod_{a=1}^4\frac{d^3\bq_a}{(2\pi)^3}\tW(\bq_a)\right]\nonumber\\
&&\times (2\pi)^3\delta_D^3(\bq_{1234})
T_\delta(\bk+\bq_1,-\bk+\bq_2,\bk'+\bq_3,-\bk'+\bq_4),
\end{eqnarray}
where $T_\delta$ is the mass trispectrum and the notation $\bq_{1\dots
n}=\bq_1+\cdots+\bq_n$.  The convolution with the window function means
that different 4-point configurations separated by less than the
Fourier width of the window function contribute to the covariance. We
call this aspect of the covariance the super sample covariance (SSC)
effect.

The trispectrum consistency introduced in \cite{TakadaHu:13} asserts
that the SSC term in the trispectrum must be consistent with the
response of the power spectrum to change in the background density:
\begin{equation}
 \bar{T}_{\delta}(\bk,-\bk+\bq_{12},\bk',-\bk'-\bq_{12})\simeq 
T_\delta(\bk,-\bk,\bk',-\bk')+\frac{\partial P_\delta(k)}{\partial
\delta_b}
\frac{\partial P_\delta(k')}{\partial \delta_b}P^L_{\delta}(q_{12}),
\end{equation}
where the mode of $\bq_{12}$ is a super-survey mode satisfying $k,k'\gg
q_{12}$, and the overbar refers to an angle average over the direction
of $\bq_{12}$. The background density $\delta_b$ is the average density
fluctuation in the survey region. Here $P^L_\delta(q)$ is is the linear
power spectrum and is designated that for this relation to be applicable
$\delta_b$ must be a mode in the linear regime, i.e. the survey scale
must be much larger than the nonlinear scale. With this consistency
prescription, the power spectrum covariance is simplified as
\begin{equation}
 C^P_{ij}=C^G_{ij}+C^{T0}_{ij}+\sigma_b^2
\frac{\partial P_\delta(k)}{\partial
\delta_b}
\frac{\partial P_\delta(k')}{\partial \delta_b},
\label{eq:pscov}
\end{equation}
where $\sigma_b^2$ is the variance of the background density in the
survey window, defined as
\begin{equation}
 \sigma_b^2\equiv \frac{1}{V_W}\int\!\frac{d^3\bq}{(2\pi)^3}
|\tW(\bq)|^2P^L_\delta(q).
\end{equation}
Here $C^{T0}_{ij}$ is the standard non-Gaussian term arising from the
mass trispectrum of sub-survey modes: $C^{T0}_{ij}=(1/V_W)\int_{|\bk|\in
k_i}\!(d^3\bk/V_{k_i}) \int_{|\bk'|\in k_j}\!(d^3\bk'/V_{k_j})
T_\delta(\bk,-\bk,\bk',-\bk')$.  The linear variance $\sigma_b$ can be
easily computed for any survey geometry, either by evaluating the above
equation directly, or using Gaussian realizations of the linear density
field.  The SSC term scales with the survey volume only through
$\sigma_b^2$ whereas the other terms scale like white noise
$1/V_W$. Thus, even if the initial density field is Gaussian, the
nonlinear structure formation induces non-Gaussian contributions to the
sample variance. In other words, the non-Gaussian sample variance
depends on the nature of large-scale structure formation that governs
how the different Fourier modes are correlated with each other via
nonlinear gravity.

\begin{figure}[t]
    \centering
    \includegraphics[width=3.5in]{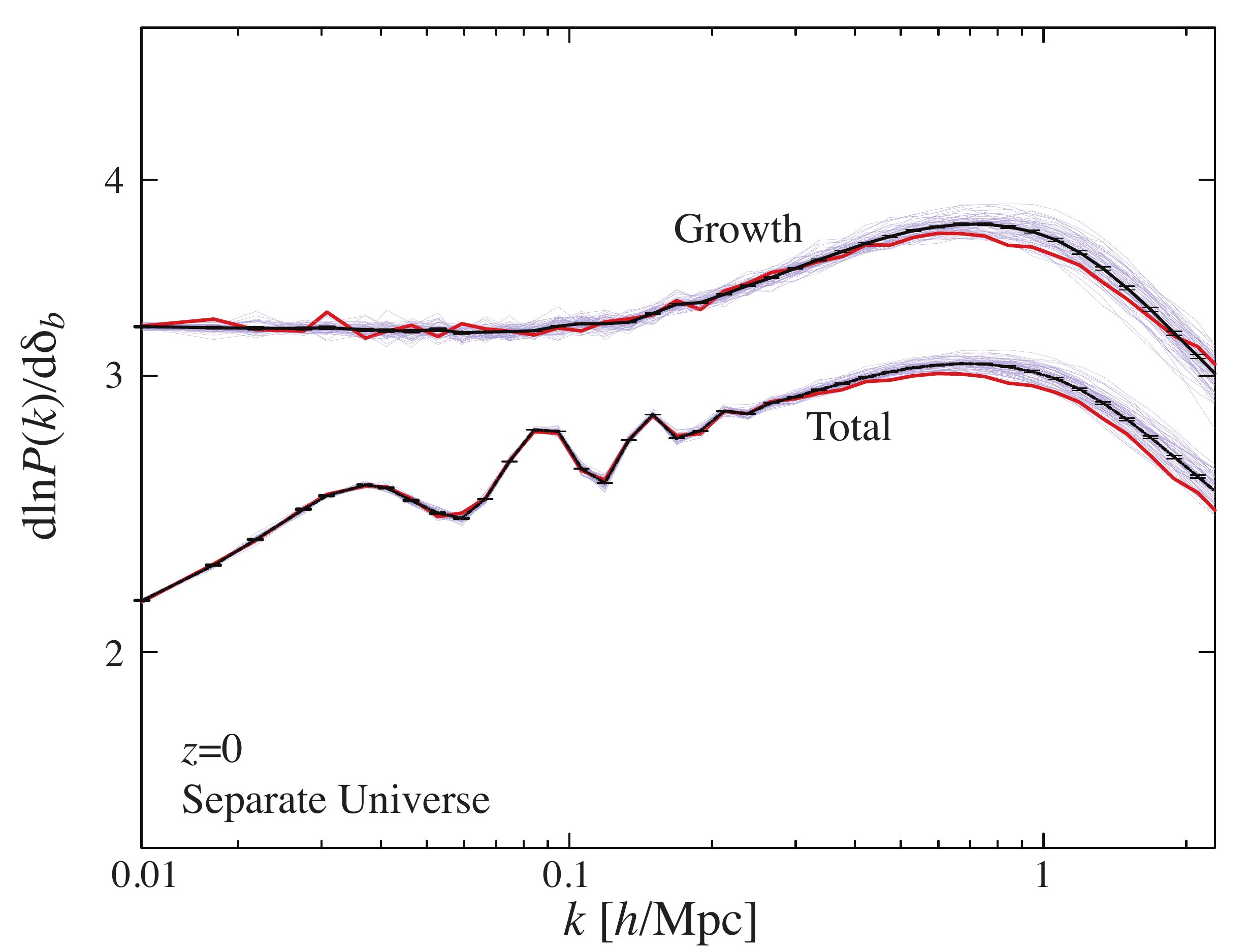}
\caption{The response of the mass power spectrum to the super-box mode
 $\delta_b$, computed using the separate universe approach. In addition
 to the fiducial run with $\delta_b=0$, we ran 64 separate universe
 pairs with $\delta_b=\pm 0.01$, where we used the same initial seeds to
 reduce the stochasticity (each simulation has a box size of
 500~$h^{-1}$Mpc and $256^3$ particles). There are two distinct effects
 of the super-box mode (here treated as a DC mode): the growth effect
 and the dilation effect (see text for details). The curve labeled by
 ``total'' is a sum of these two effects, leaving characteristic scale
 dependence in the response. The bold curve is the average of the 64
 pairs, and the thin curve is the result for each pair. The red curve is
 one particular realization. This plot is
 taken from Li \etal~(2014).  } \label{fig:dPdd_scatter}
\end{figure}
To compute the power spectrum response for a given cosmological model,
we can use the {\em separate universe approach} developed in
\cite{Lietal:14}.  In this approach the impact of the super-box mode
$\delta_b$ is absorbed into changes in the background cosmological
parameters in a finite-volume simulation with periodic boundary
condition:
\begin{eqnarray}
&&\frac{\delta a}{a}=-\frac{1}{3}\delta_b,\hspace{1em}
%\nonumber\\
%&&
\frac{\delta h}{h}=-\frac{5\Omega_m}{6}\frac{\delta_b}{D},\hspace{1em}
%\nonumber\\
%&&
\frac{\delta \Omega_m}{\Omega_m}=\frac{\delta
 \Omega_\Lambda}{\Omega_{\Lambda}}=\frac{\delta
 \Omega_K}{1-\Omega_K}=-\frac{\delta h}{h}.
\end{eqnarray}
Here $D$ is the linear growth rate and we have introduced the notations
such as $\delta h/h=(H_{0W}-H_0)/H_0$, where $H_{0W}$ denotes the
parameter in a separate universe (a finite volume region at the fixed
$\delta_b$). Our convention is to set the scale factor of the separate
universe $a_W$ to agree with the global one at high redshift:
$\lim_{a\rightarrow 0} a_W(a,\delta_b)=a$.  Since the linear background
density $\delta_b$ evolves with $D$, so $\delta_b/D=\delta_{b0}/D_0$;
the relations about cosmological parameters hold independently of the
redshift at which $\delta_b$ and $D$ are defined. Thus, even if the
global universe has a flat geometry, $\Omega_K=0$, the separate universe
with non-zero $\delta_b$ is realized as a non-zero curvature universe,
$\Omega_{KW}\ne 0$. Because this is the curvature effect, time evolution
of all the sub-box Fourier modes is affected by the super-box mode, due
to the modified expansion history.

Fig.~\ref{fig:dPdd_scatter} shows the power spectrum response computed
using the separate universe approach.  There are two distinct
contributions to the power spectrum response. First, the presence of
super-survey mode modifies the growth of sub-volume modes via the mode
coupling in nonlinear structure formation. If the survey region is
embedded in a coherently overdense region, i.e. $\delta_b>0$, the growth
of sub-volume modes is enhanced. We call this effect ``growth''. Second,
the super-survey mode causes remapping of physical and comoving length
scales. An overdense region expands less quickly than in the global
universe. We call this effect ``dilation'' as it changes the comoving
scale corresponding to features in the power spectrum. The figure shows
that, in the total, these two effects partially cancel, leaving a
characteristic scale-dependence in the response. 
We also note that, if
we use the
halo model to estimate the power spectrum response by directly
computing the windowed trispectrum, $\bar{T}_W(k_i,k_j)$, the 
analytical prediction gives about 10\%-level agreement with the separate
universe result over the range of $k$ shown in
Fig.~\ref{fig:dPdd_scatter}.

\begin{figure}
    \centering
    \includegraphics[width=3.in]{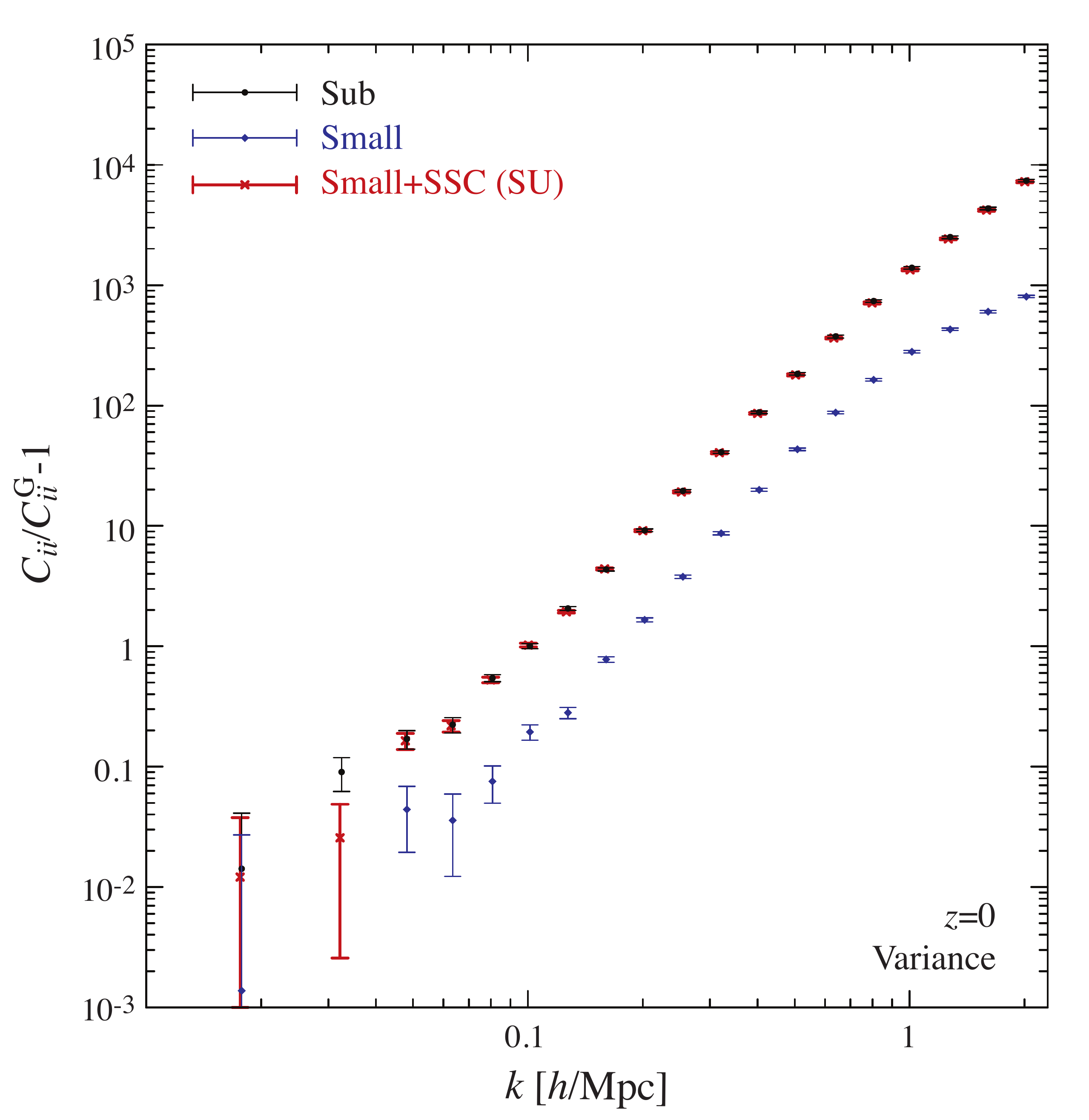}
\caption{Diagonal elements of the mass power spectrum covariance,
$C_{ij}$, relative to the Gaussian term $C_{ii}^{G}$ at $z=0$.  The
result denoted as ``Sub'' is the covariance estimated from subvolumes of
7 large-volume simulations; each of the simulations has a $4~{\rm
Gpc}/h$ box size and is divided into $8^3=512$ subvolumes of size
500~${\rm Mpc}/h$ each (3584 subvolumes in total). Thus each subvolume
includes the super-box mode effects.  The result ``Small'' is the
covariance estimated from small-box simulations of 500~${\rm Mpc}/h$
each, with periodic boundary conditions. The result ``Small+SSC'' shows
the covariance computed by adding the SSC effect, calibrated based on
the separate universe approach in the previous figure, to the small-bx
variance. The ``Small+SSC'' result is in nice agreement with the ``Sub''
result to within the bootstrap errors.
%The SSC effect causes the subboxes of the large volume
%simulations to have up to an order of magnitude higher variance than
%found in small periodic boxes of the same volume.  Adding the SSC
%separate universe (SU) response to a background mode to the the small
%box variance models the effect to within the bootstrap errors of the
%simulation suites. 
Note that bootstrap errors between bins 
 are highly correlated. This plot is taken from Li \etal~(2014).
} \label{fig:variance}
\end{figure}
Upcoming wide-area galaxy surveys require an accurate estimation of the
power spectrum covariance or more generally the covariance of any
large-scale structure probes. This is indeed computationally
challenging.
%Hence, a calibration of the power spectrum
%covariance, more generally the covariance of any large-scale structure
%probe, is computationally challenging. 
With the unified theory of the covariance, Eq.~(\ref{eq:pscov}), we can
propose a way of calibrating the power spectrum covariance at
computationally reasonable expense. To compute the standard part, the
Gaussian piece and the trispectrum piece of sub-volume modes, we can use
mock catalogs of a galaxy survey, based on N-body simulations of small
boxes. To compute the SSC effect, we can use the separate universe
simulations for the fiducial cosmology. In doing this, we can properly
take into account the survey window to compute the linear variance,
$\sigma_b^2$. This method does not require huge-volume simulations whose
size is designed to be well larger than the size of survey volume in
order to include the super-survey effects. 

In Fig.~\ref{fig:variance}, we indeed show that the above method combining the
small-box simulations and the separate universe simulations well
reproduces the covariance matrix from the large-volume simulations. The
figure also shows that the SSC effect boosts the covariance amplitude by
up to an order of magnitude over the range of wavenumbers we
consider. Hence the SSC effect is the dominant source of the sample
variance. This results hold for any size of survey volumes relevant for
upcoming galaxy surveys (see Fig.~1 in Takada \& Hu 2013).

If the power spectrum needs to be estimated with respect to the local
average density within the finite-volume survey region, which is the
case for the galaxy power spectrum (Takada et al. 2014), the power
spectrum response is modified as
\begin{equation}
 \frac{\partial \ln P^W(k)}{\partial \delta_b}=\frac{\partial \ln P(k)}{\partial
 \delta_b}-2,
\end{equation}
where $P^W(k)=P(k)/(1+\delta_b)^2$ is the power spectrum with respect to
the local average density. The SSC effect is reduced in the covariance
of the local power spectrum, but still gives a dominant contribution in
the nonlinear regime (Li \etal~2014).

\begin{figure}
    \centering
    \includegraphics[width=0.48\textwidth]{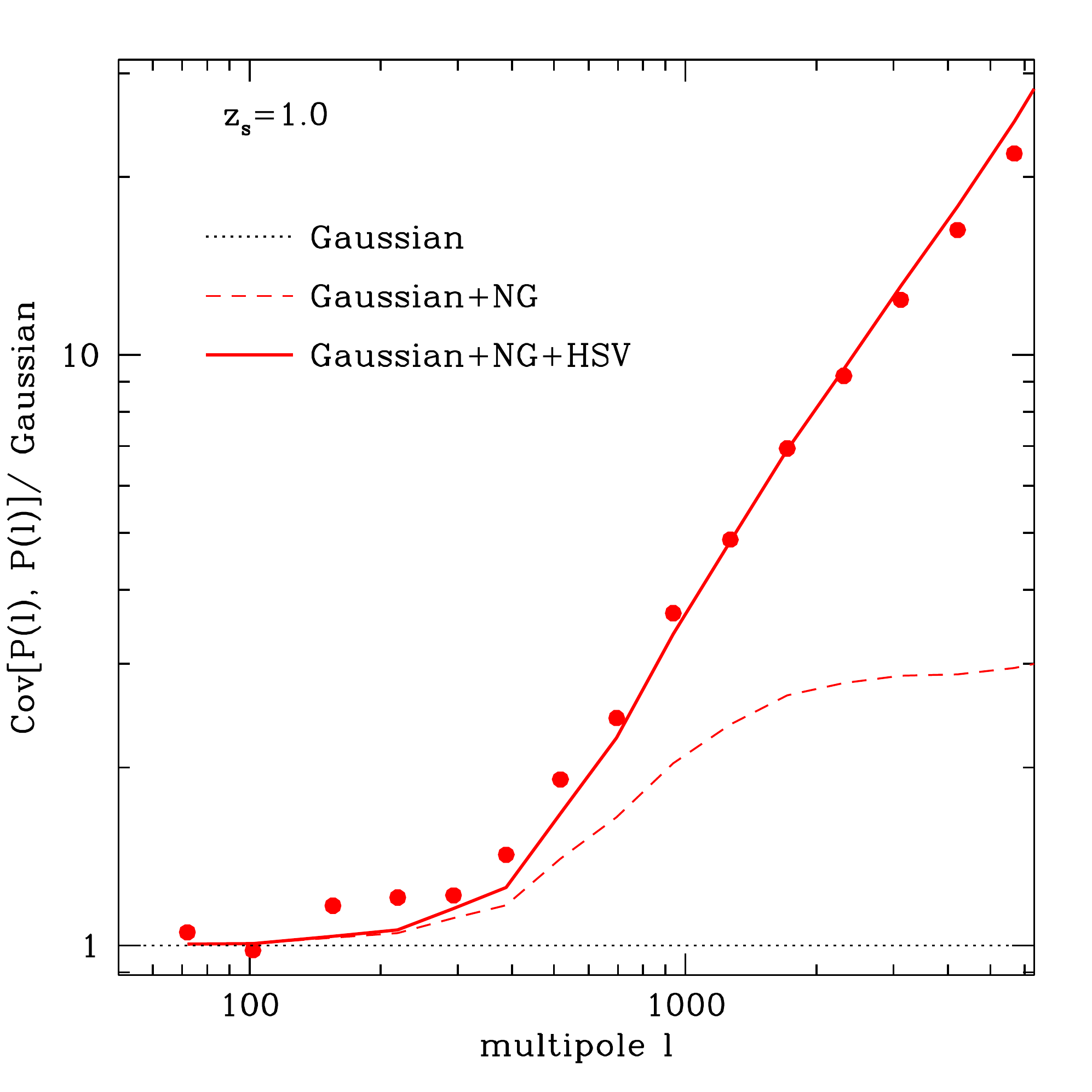}
    \includegraphics[width=0.50\textwidth]{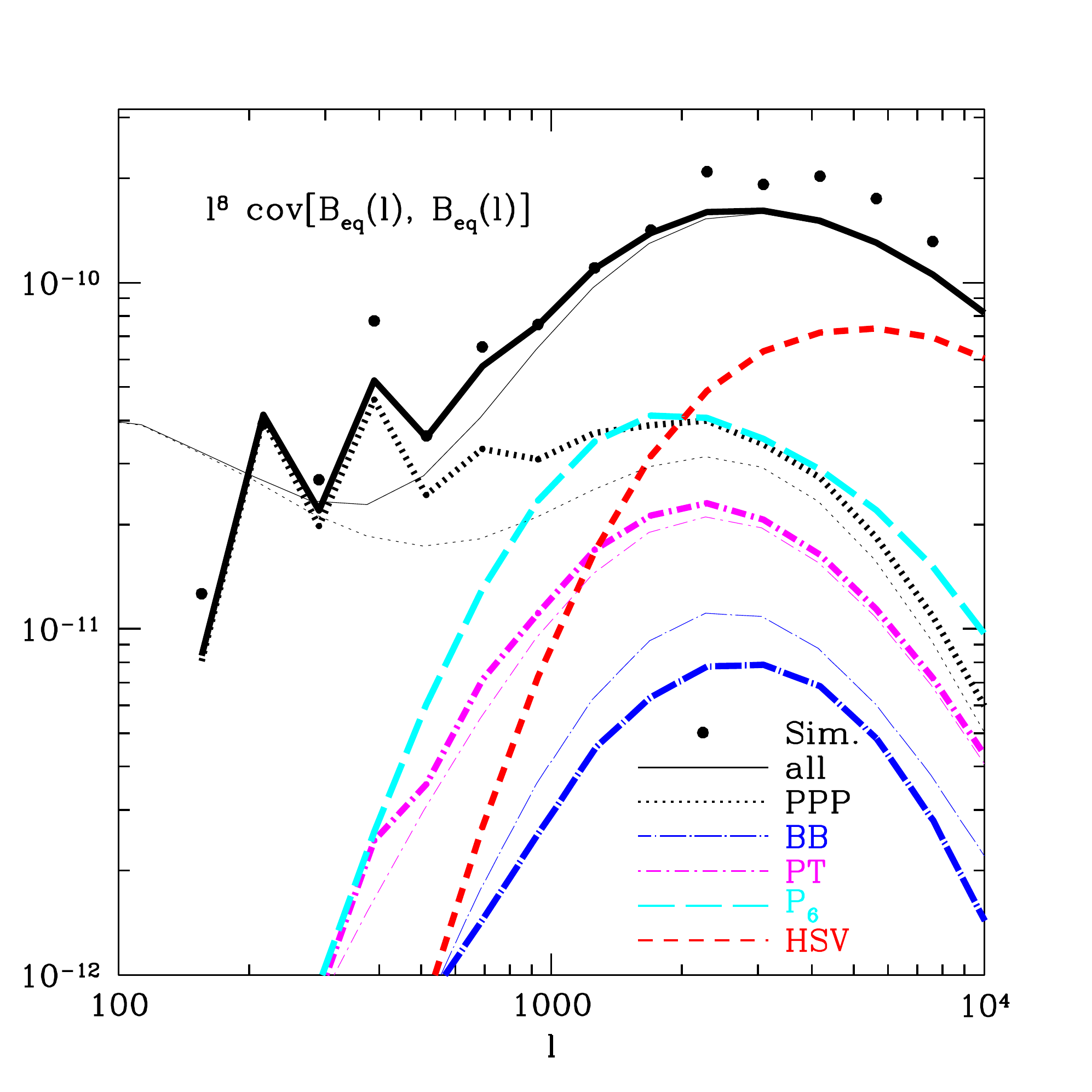}
\caption{{\em Left panel}: Diagonal elements of the weak lensing power
spectrum covariance, relative to the Gaussian covariance.  Note that we
here ignored shape noise contribution.  The circle points show the
results measured from 1000 realizations of the ray-tracing simulations
for the $\Lambda$CDM model, each of which is for source redshift $z_s=1$
and has an area of 25 sq. degrees. Note that we used the
logarithimcally-spacing bins of $\Delta \ln\ell=0.3$.  The solid or
dashed curves are the analytical predictions with or without the HSV
contribution, which is the small-scale, non-linear version of the SSC
effect (see text for details). {\em Right panel}: Diagonal elements of
the weak lensing bispectrum for equilateral triangle configurations
against the triangle side length. The data points are the results
measured from the 1000 ray-tracing simulations.  The bold solid curve is
the total power including the HSV effect, and fairly well reproduces the
simulation results.  The other curves are each contribution that arises
from the HSV effect or the 2-, 3-, 4- and 6-point correlation functions
as indicated.  \label{fig:wl_hsv}}
\end{figure}
The small-scale, nonlinear version of the SSC effect can also be
realized within the framework of the halo model approach -- the halo
sample variance (HSV) (\cite[Hu \& Kravtsov 03; Takada \& Bridle 2007; Takada \& Jain 2009;
Sato \etal~ 2009; Kayo \etal~ 2013; Takada \& Hu 2013; Takada \& Spergel
2014; Schaan et al
2014]{Hamiltonetal:06,HuKravtsov:03,TakadaBridle:07,TakadaJain:09,Satoetal:09,Kayoetal:13,TakadaHu:13,TakadaSpergel:14,Schaanetal:14}). In
the halo model formulation, the 1-halo term of the mass power spectrum,
which describes correlations between dark matter in the same halo, is
expressed as
\begin{equation}
 P^{1h}_{\delta}(k)=\int\!\!dM\frac{dn}{dM}\left(\frac{M}{\bar{\rho}_m}\right)^2|\tilde{u}_M(k)|^2,
\label{eq:p1h}
\end{equation}
where $dn/dM$ is the halo mass function in the mass range $[M,M+dM]$,
$\bar{\rho}_m$ is the cosmic mean mass density, and $\tilde{u}_M(k)$ is
the Fourier transform of the mass density profile of halos of mass
$M$. However, the above equation is correct only in an ensemble average
sense. For a finite-volume survey, the coherent density fluctuation
across the survey window, $\delta_b$, would change the abundance of
halos in the survey region along the peak-background splitting theory:
\begin{equation}
\left.\frac{dn}{dM}\right|_{\delta_b}=\frac{dn}{dM}\left[1+b(M)\delta_b+\cdots\right],
\label{eq:dn_db}
\end{equation}
where the notation $|_{\delta_b}$ denotes the average over the
realizations of different sub-survey modes at fixed $\delta_b$, and
$b(M)$ is the halo bias. Thus, e.g., if the survey region is in a
coherent over-density region, it enhances the number of halos on
average. 

By inserting Eqs.~(\ref{eq:p1h}) and (\ref{eq:dn_db}) into the
covariance formula (Eq.~\ref{eq:pscov}) via the trispectrum consistency
relation, we find that the change in the halo mass function via the
super-survey modes causes co-variant scatters in the power spectrum
estimation:
\begin{eqnarray}
 C^{\rm HSV}_{ij}&=&\sigma_b^2\frac{\partial P^{1h}(k_i)}{\partial
  \delta_b}
\frac{\partial P^{1h}(k_j)}{\partial \delta_b}
\nonumber\\
&=&\sigma_b^2
\left[
\int\!\!dM\frac{dn}{dM}b(M)|\tilde{u}_M(k_i)|^2
\right]
\left[
\int\!\!dM'\frac{dn}{dM'}b(M')|\tilde{u}_{M'}(k_j)|^2
\right],
\end{eqnarray}
where we have assumed that the super-survey modes do not affect the halo
mass profile. We found that the HSV effect gives a dominant contribution
of the SSC effect in the power spectrum covariance at $k\simgt $ a few
$0.1~h/{\rm Mpc}$, fairly well reproducing the separate universe
simulation results at the scales (\cite[see Fig.~2 Takada \& Hu 2013 or
Fig.~1 in Li \etal~ 2014]{TakadaHu:13,Lietal:14}).

\subsection{Information content of lensing power spectrum and 
bispectrum}

Similarly to the mass power spectrum covariance, we can compute the
covariance matrices for the weak lensing power spectrum and bispectrum
(\cite[Takada \& Bridle 2007; Takada \& Jain 2009; Sato \etal~ 2009;
Kayo \etal~ 2013; Takada \& Spergel 2014; Schaan \etal~%
2014]{TakadaBridle:07,TakadaJain:09,Satoetal:09,Kayoetal:13,TakadaSpergel:14,Schaanetal:14}).
The left panel of Fig.~\ref{fig:wl_hsv} shows diagonal elements of the
weak lensing power spectrum covariance relative to the Gaussian
covariance, measured from the ray-tracing simulations that are built
using a suite of N-body simulations for the WMAP $\Lambda$CDM model
(\cite[Sato \etal~ 2009]{Satoetal:09}). We used 1000 realizations for
source redshift $z_s=1$ each of which has an area of 25 sq. degrees
corresponding to the fundamental Fourier mode $l_f\simeq 72$. The
ray-tracing simulations were done in a light cone volume with an
observer's position being its cone vertex, and therefore include
contributions from N-body Fourier modes with length scales greater than
the light-cone volume at each lens redshift (see Fig.~1 in Sato
\etal~2009). Thus the simulations are suitable to study the SSC
effect. As can be found from the figure, the weak lensing power spectrum
covariance shows significant non-Gaussian errors at $\ell \simgt $ a few
hundreds. The solid curve denotes the analytical prediction including
the HSV effect, showing remarkably nice agreement with the ray-tracing
simulation result. We note that the HSV effect causes highly correlated
scatters between different multipoles.  If we ignore the HSV effect,
i.e. include the standard non-Gaussian error alone arising from the
lensing trispectrum of sub-survey modes, the model prediction
significantly underestimates the total power.

Similarly the right panel of Fig.~\ref{fig:wl_hsv} shows the results for
the bispectrum covariance matrix.  Here we consider the equilateral
triangle configurations. The HSV effect
gives a dominant contribution to the total power of the covariance
matrix at $\ell \simgt 1000$, compared to other terms up to the 6-point
correlation function.  If we include the HSV contribution, the
analytical model gives a 10-20\% level agreement with the simulation
results.

\begin{figure*}
\begin{center}
 \includegraphics[width=0.6\textwidth]{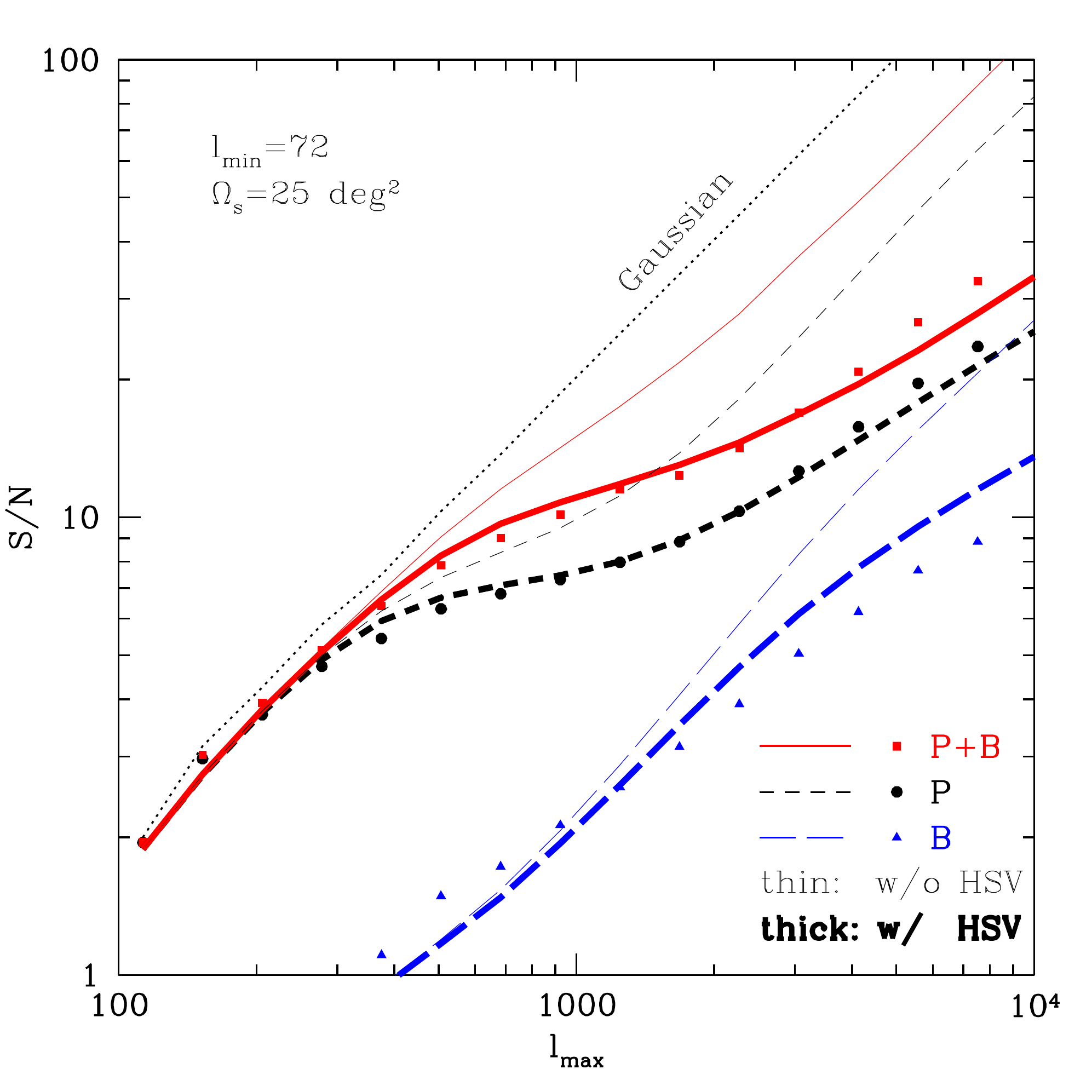} 
\caption{
Cumulative signal-to-noise ratios ($S/N$) for the power spectrum ($P$),
the bispectrum ($B$) and the joint measurement ($P+B$) for a survey area
of 25 deg$^2$ and source redshift $z_s=1$.  
%They are shown as functions
%of the maximum multipole $l_{\rm max}$, where the power spectrum and/or
%bispectrum information are included over $l_{\rm min}\le l \le l_{\rm
%max}$ (see Eqs.~\ref{eq:snps1}, \ref{eq:snbs1} and \ref{eq:snp+b}).  
%The
%minimum multipole is set to $l_{\rm min}=72$. 
%We do not include the
%shape noise contamination here -- it is shown in the next figure.  
The circle, triangle and square symbols are the simulation results for
$P$, $B$ and $P+B$ measurements, respectively, computed from the 1000
realizations. To account for the full bispectrum information, we
included the bispectra of {\rm all}-available triangle configurations
from the multipole range, up to 204 triangles for $l_{\rm max}=8745$.
The thick short-dashed, long-dashed and solid curves are the
corresponding halo model predictions.  The corresponding thin curves are
the results without the HSV contributions.
% with/without the HSV terms.
For comparison, the dotted curve shows the $S/N$ for the power spectrum
for the Gaussian field, which the primordial density field should have
contained.  The Gaussian information follows a simple scaling as 
$S/N|_{\rm Gaussian}\propto l_{\rm max}\Omega_s^{1/2}$, where $\Omega_s$
 is the survey area.
This plot is from Kayo \etal~(2013).
\label{fig:sn}} 
\end{center}
\end{figure*}
Once the covariance matrices for the weak lensing power spectrum and
bispectrum are computed, we can address the information content carried
by the weak lensing correlation functions. 
%Since large-scale structure
%originates from the initial Gaussian field, we can address the
%fundamental question: whether or not we can recover the Gaussian
%information from measurements of large-scale structure observables. Here
%we address this question for the weak lensing observables.
For a given range of multipoles, the cumulative signal-to-noise ratio or
the information content for the power spectrum measurement is defined as
\begin{equation}
 \left(\frac{S}{N}\right)^2_{P}\equiv 
\sum_{l_{\rm min}\le l,l'\le l_{\rm max}}P_\kappa(l)[{\bf C}^{-1}]_{ll'}P_\kappa(l'),
\end{equation}
where the summation runs over all multipole bins in the range $l_{\rm
min}\le l\le l_{\rm max}$ and ${\bf C}^{-1}$ is the inverse of the
covariance matrix.  The inverse of $S/N$ is equivalent to a precision of
measuring the logarithmic amplitude of the power spectrum up to a given
maximum multipole $l_{\rm max}$, assuming that the shape of the power
spectrum is perfectly known. Similarly we can define the $S/N$ values
for the bispectrum measurement and for a joint measurement of the power
spectrum and the bispectrum. For the latter case, we need to properly
take into account their cross-covariance.

Fig.~\ref{fig:sn} shows the expected $S/N$ for measurements of the weak
lensing power spectrum and bispectrum for a survey area of 25 square
degrees (i.e. the area of the ray-tracing simulation), as a function of
the maximum multipole $l_{\rm max}$ up to which the power spectrum
and/or bispectrum information are included. The minimum multipole is
fixed to $l_{\rm min}=72$. Roughly speaking the $S/N$ value scales
with survey area as $S/N\propto \Omega_s^{1/2}$ (exactly speaking the
scaling does not hold due to the different dependence of the SSC effect
on survey area).  The circle, triangle and square symbols are the
simulation results for the power spectrum, the bispectrum and the joint
measurement, respectively. For the bispectrum measurement we included
the bispectra of all triangle configurations available from the
multipole range, and considered up to 204 triangles for $l_{\rm
max}=8745$.  The thick/thin short-dashed, long-dashed and solid curves
are the analytical predictions with/without the HSV effects. First, the
figure clearly shows that the lensing bispectra add new information
content to the power spectrum measurement. To be more quantitative,
adding the bispectrum measurement increases the $S/N$ by about 50 per
cent for $l_{\rm max}\simeq 10^3$ compared to the power spectrum
measurement alone.  This improvement is equivalent to about 2.3 larger
survey area for the power spectrum measurement alone; that is, the same
data sets can be used to obtain the additional information, if the
bispectrum measurement is combined with the power spectrum
measurement. Secondly, the analytical predictions are in nice agreement
with the simulation results. Note that the total $S/N$ for the joint
measurement $(P + B)$ is close to the linear sum of the $S/N$ values,
not the sum of their squared values $(S/N)^2$, due to the significant
cross-covariance.  If ignoring the cross-covariance, adding the
bispectrum measurement does not much improve the $S/N$ (only by 5 per
cent or so).

The top, dotted lines shows the information content for a Gaussian
field, which the initial density field of large-scale structure should
have contained -- therefore can be considered as a maximum information
content we could ultimately extract. The Gaussian information content
depends only on the number of Fourier modes available from the range of
multipoles up to $l_{\rm max}$: it has a simple scaling given by
$\left.S/N\right|_{\rm Gaussian}\propto \Omega_{s}^{1/2}l_{\rm max}$.
The figure shows that the joint measurement can recover only
about 50\% or less of the Gaussian information at $l_{\rm max}\simgt
1000$. The information loss is mainly due to the HSV effect, as can be
found from the thin curves. If we ignore the HSV effect, the joint
measurement recovers about 70\% of the Gaussian information at $l_{\rm
max}\simeq 1000$. These results imply that further higher-order
functions such as the 4-point function may be important and add the
information (\cite[see Seo \etal~2011]{Seoetal:11}).  Or some of the
initial Gaussian information is lost or cannot be recovered due to the
nonlinear clustering. This is not yet known, and still an open question.

\subsection{Weak lensing tomography}
\begin{figure}
\centering
\includegraphics[width=0.49\textwidth]{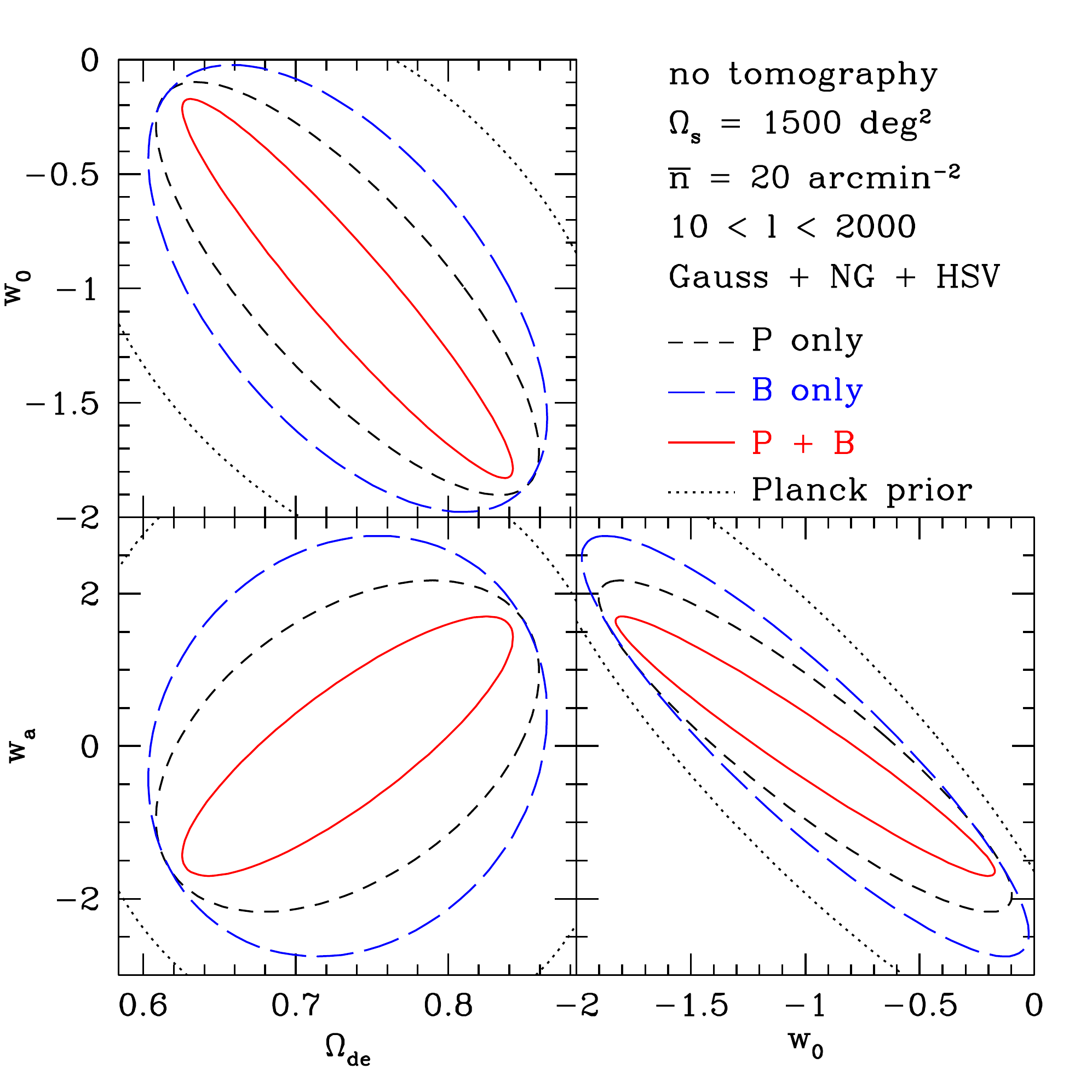}
\includegraphics[width=0.49\textwidth]{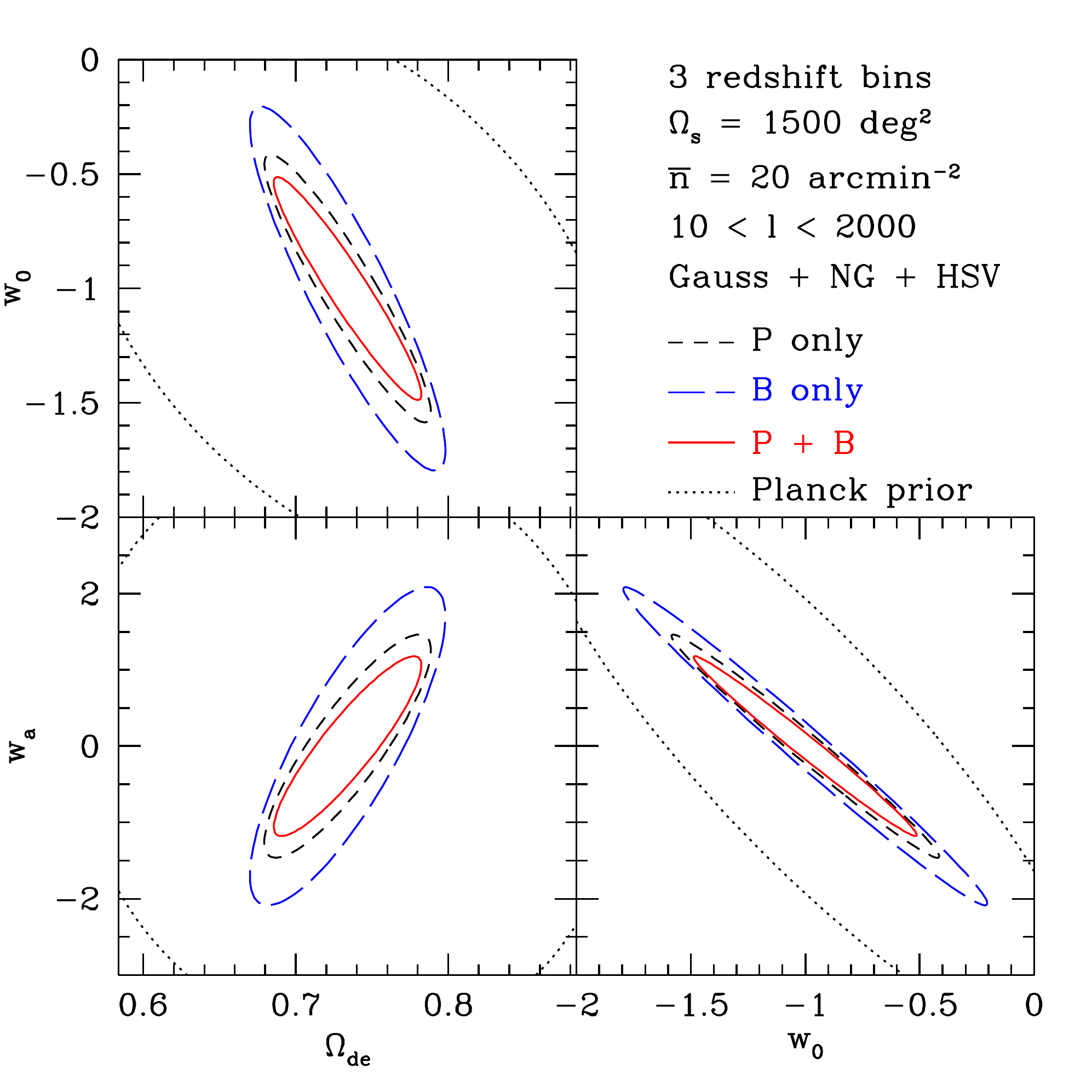}
\caption{Expected accuracies of dark energy parameters ($\Omega_{\rm de},
 w_0, w_a$) for a galaxy survey that 
resemble the Subaru Hyper
 Suprime-Cam survey ($\Omega_s=1500$ sq. degrees,
 $\bar{n}_g=20$~arcmin$^{-2}$ and $\sigma_\epsilon=0.22$). For this plot
 we included the shape noise contribution to the covariance.
The error
 ellipses include marginalization over other parameters, and we included
 the CMB information expected for the Planck experiment. We here show
 the parameter forecasts for the power spectrum information alone ($P$),
 the bispectrum alone ($B$) and the joint information ($P+B$) as in the
 previous plot. The left and right panels show the results for no
 tomography (single redshift bin) and three-redshift tomography case. As
 in the previous figure, we took into account the full covariance
 between the observables including the HSV effects. For the tomography
 case, we included 6525 different bispectra in the multipole range
 $10\le l\le 2000$ that are constructed from all combinations of
 different redshift bins and triangle configurations.
This plot is from Kayo \& Takada (2014).
} \label{fig:deparas}
\end{figure}

Adding redshift information to the weak lensing correlation functions
greatly improves the cosmological sensitivity -- the so-called weak
lensing tomography (\cite[Takada \& Jain 2004]{TakadaJain:04}). However,
to realize the genuine power of the weak lensing tomography, we need to
consider all the spectra available from all possible combinations of
different redshift bins and multipole bins. For the bispectrum case,
adding the lensing tomography easily leads to more than 1000 bispectra,
and this is even worse for the higher-order correlations.  Hence, an
accurate calibration of the covariance matrix for the lensing tomography
would require a huge number of independent ray-tracing simulations,
e.g., a factor 10 more realizations than the number of bispectra to
achieve about 10\% accuracy.  This is computationally very challenging,
and is even impossible if we need to compute the covariance as a function
of cosmological models. So again a hybrid method combining small-box
simulations, separate-universe simulations, and the analytical model
would be useful and
tractable in practice. 

In Kayo \& Takada (2013), we used the halo model approach, which gives a
fairly good agreement with the simulation results for no tomography case
as shown in Fig.~\ref{fig:sn}, in order to estimate the cosmological
power of the weak lensing bispectrum tomography. Fig.~\ref{fig:deparas}
shows expected accuracies of dark energy parameters assuming survey
parameters that resemble the Subaru Hyper Suprime-Cam, characterized by
$\Omega_{\rm s}=1500$ sq. degrees, $n_{g}=20$ arcmin$^{-2}$ and
$\sigma_\epsilon=0.22$ for the survey area, the mean number density of
source galaxies and the rms intrinsic ellipticities per component,
respectively. 
%The figure clearly shows that, even if including the three
%redshift-bin tomography information, 
The bispectrum further adds the information to improve the parameter
constraints compared to the power spectrum alone. To be more precise,
for the three redshift bin case (the right panel), the joint measurement
leads to about 60\% improvement in the dark energy figure-of-merit (FoM)
that is defined by ${\rm FoM}=1/[\sigma(w_0)\sigma(w_a)]$. Again this is
equivalent to about 60\% larger survey area for the power spectrum
tomography alone.  In this case we considered 6525 triangle
configurations, and we take into account the non-Gaussian correlations
between the different spectra including the HSV effects.

\section{Discussion}

Can we recover the initial Gaussian field from observables of the
present-day large-scale structure? This is an unresolved, open
question. In this paper we discussed the example of the weak lensing
field that is the line-of-sight projection of the three-dimensional mass
density field in large-scale structure. We showed that the information
content inherent in the power spectrum, which is the two-point
correlation function in Fourier space, is smaller than the Gaussian
information by more than a factor of 2 at $l\simgt 1000$. We showed that
the bispectrum, which is the three-point correlation function, does add
the information to the power spectrum, but the combined information does
not fully recover the Gaussian information -- still only 50\% of the
Gaussian information at $l_{\rm max}\sim 1000$. In order to derive this
conclusion, we included all the two- and three-point level information
in a sense that we included the bispectra of all available triangle
configurations for a given range of multipoles as well as properly took
into account the auto- and cross-covariances between the two- and
three-point correlation functions.  This implies that
the higher-order correlation functions are further needed to recover the
Gaussian information. 
Or our result implies a
limitation of the information recovery; some of the initial memory is
lost due to the strong mode coupling in the deeply nonlinear regime. 
Alternatively, the nonlinear mapping of the
cosmological field, such as the log-normal mapping (Seo \etal~2012),
might be a more practically useful way. 
%Again these are not yet resolved and require a further careful
%study.

However, this conclusion is a bit misleading. Most of the information
loss is caused mainly by the super-sample covariance. As we showed, the
super-survey effects are parameterized mainly by the average density
fluctuation in the survey volume, $\delta_b$, on each realization
basis. Hence, by treating $\delta_b$ as an additional parameter together
with cosmological parameters, we may be able to calibrate most of the
super-sample variance effects in the correlation measurements
(\cite[e.g., see Takada \& Spergel 2013; Schaan \etal~ 2014; Li \etal~%
in preparation]{TakadaSpergel:13,Schaanetal:14,Lietal:14b}). We can even
treat the super-survey mode as ``signal'' and then estimate its value
for a given survey volume by fitting the measurements with the
model. This is an interesting possibility, and needs to be further
explored. A physical understanding of the super-survey effects is also
important to explore an optimal survey design that allows an efficient
operation of massive cosmological surveys (Takahashi \etal~2014).

\smallskip{\em Acknowledgments.--} MT was supported by World Premier
International Research Center Initiative (WPI Initiative), MEXT, Japan,
by the FIRST program ``Subaru Measurements of Images and Redshifts
(SuMIRe)'', CSTP, Japan, and by Grant-in-Aid for Scientific Research
from the JSPS Promotion of Science (Nos. 23340061 and 26610058).

\end{document}